\documentclass[journal=jacsat,manuscript=article]{achemso}

\usepackage{times}
\usepackage{color}
\usepackage{graphicx}
\usepackage{amsmath,amsbsy,amsfonts,mathrsfs}

\newcommand{\be}{\begin{equation}}
\newcommand{\ee}{\end{equation}}
\newcommand{\bea}{\begin{eqnarray}}
\newcommand{\eea}{\end{eqnarray}}
\newcommand{\bg}{\begin{figure}}
\newcommand{\eg}{\end{figure}}
\newcommand{\bi}{\begin{itemize}}
\newcommand{\ei}{\end{itemize}}
\author{Jos\'e I. Mart\1nez}
\email{joseignacio.martinez@uam.es}
\affiliation[Universidad Aut\'onoma de Madrid]{Departamento de F\1sica Te\'orica de la Materia Condensada, Universidad
Aut\'onoma de Madrid, E-28049 Madrid, Spain}
\author{Federico Calle-Vallejo}
\affiliation[Leiden University]{Leiden Institute of Chemistry, Leiden University, PO box 9502, 2300 RA Leiden, The
Netherlands}
\author{Clifford M. Krowne}
\affiliation[Naval Research Laboratory]{Microwave Technology Branch, Electronics Science \& Technology
Division, Naval Research Laboratory, Washington, DC 20375, United States of
America}
\author{Julio A. Alonso}
\affiliation[Universidad de Valladolid]{Departamento de F\1sica Te\'orica, At\'omica y \'Optica, Facultad
de Ciencias, Universidad de Valladolid, ES-47005 Valladolid, Spain}

\title[First-principles Structural and Electronic Characterization of Ordered SiO$_2$ Nanowires]
{First-principles Structural and Electronic Characterization of Ordered SiO$_2$ Nanowires}

\begin{document}

\newpage

\begin{abstract}
Density functional theory and molecular dynamics simulations have been used to optimize the structure of nanowires of
SiO$_2$. The starting structures were based on $\beta$-cristobalite, ortho-tridymite, $\beta$-tridymite, and rutile
crystals. The analysis of the electronic structure has been validated by many-body perturbation calculations using the 
G$_0$W$_0$ and GW + Bethe-Salpeter equation approximations, in order to account for quasi-particle and excitonic effects. 
The calculations indicate that many of these nanowires have semiconducting character, while the corresponding bulk solids 
are insulators. In the case of thick rutile-like nanowires we found the gap congested with surface states. Electronic 
charge is transferred from silicon atoms to oxygen atoms, and the bonding in the nanowires is partly ionic and partly 
covalent. The magnitude of the band gap can be engineered by changing the structure and the thickness of the nanowires.
\end{abstract}

\noindent Keywords: \textit{density functional theory; silicon dioxide; nanocables; semiconductor; surface states; many-body corrections.}


\section{INTRODUCTION}

Metallic oxide uses, including as catalysts, and in \emph{state of the art} batteries, fuel cells, solar cells, and supercapacitors for electronics~\cite{1,2,3,4,5,6,7,7a},
is recently arising as a burgeoning area, especially for nanostructures, like nanoclusters, nanotubes, nanowires and nanocables~\cite{8,9}. Recently, it has been shown that
a metallic oxide combination, RuO$_2$/SiO$_2$, consisting of a transparent conductive platinum group metal oxide and a dielectric insulating silica material, can be made in
a cylindrical layered fashion~\cite{10}. The RuO$_2$ part forms a thin shell around the inner solid core nanowire (or nanorod) of SiO$_2$. Because Ruthenia is such an
extraordinary material for charge storage - the highest known - being able to use cores of an inexpensive material to act as its scaffolding becomes an ideal way to use
the expensive metal ruthenium.

Work on determining theoretically the intrinsic quantum conductances and capacitances~\cite{11}, as well as the junction capacitances~\cite{12}, of ruthenium oxide nanowires
and nanocables, followed the earlier effort in producing conductive Ruthenia nanoskins on insulating silica paper~\cite{13}. The essential purpose of those studies was to
investigate charge storage for chemical and electronic applications. What must be done now is to completely characterize the underlying SiO$_2$ nanowires, since their properties
may affect the overlying RuO$_2$ material. Thick SiO$_2$ nanowires have been synthesized~\cite{14,15}. However, to our knowledge, few theoretical studies of SiO$_2$ nanowires
have been performed, and only for very thin nanowires~\cite{16,17}. We here present density functional calculations of the atomic and electronic structure of SiO$_2$ nanowires
as a function of nanowire thickness. An analysis of the electronic density of states, electrical character, electron density, charge transfer and chemical bonding is presented.

\begin{figure}
\centerline{\includegraphics[width=12cm]{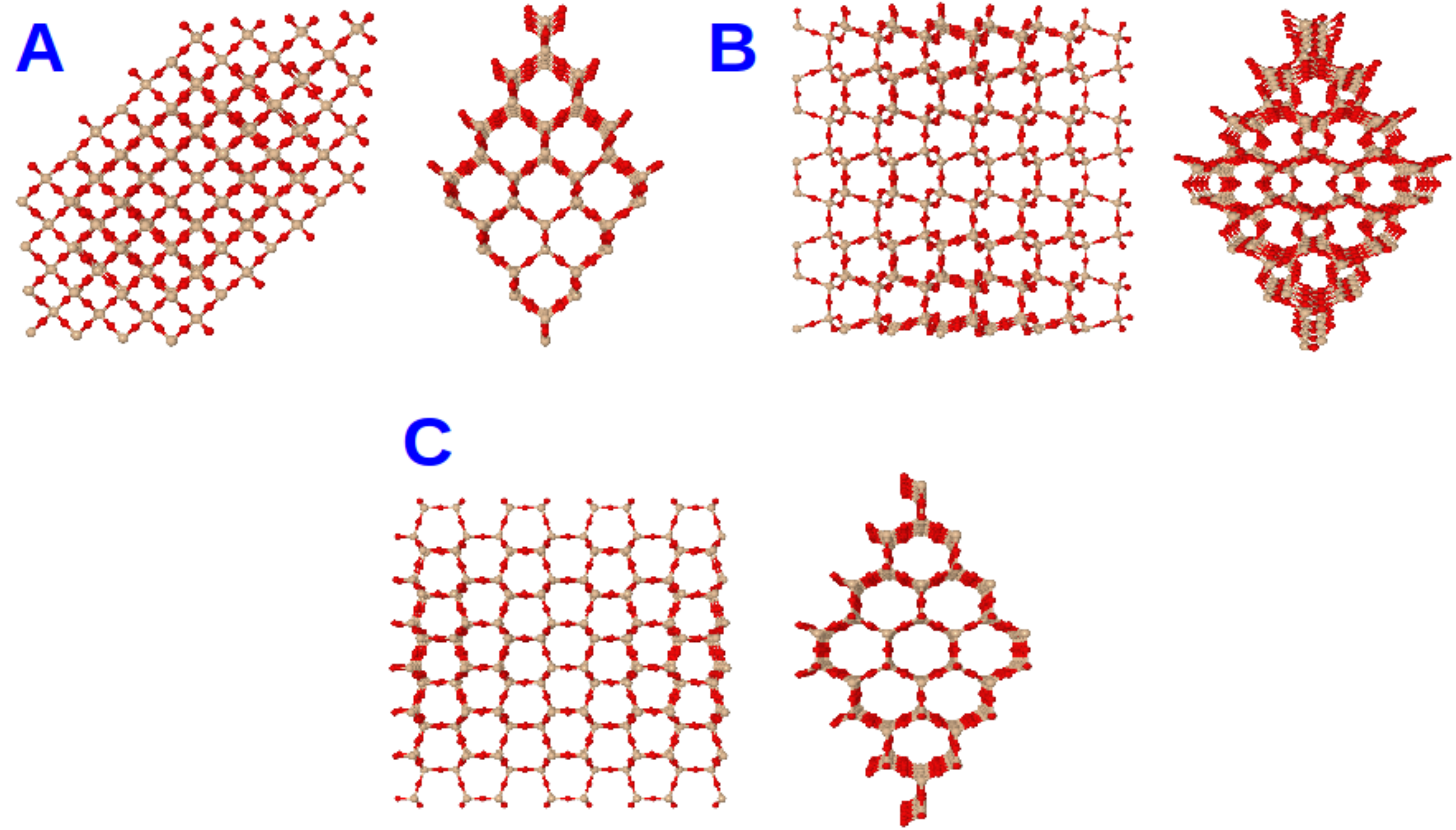}}
\smallskip \caption{(Color online) Two views of the crystalline structure of some SiO$_2$ bulk-phases:
A) $\beta$-cristobalite~\cite{27}; B) ortho-tridymite~\cite{28}; C) $\beta$-tridymite~\cite{29}.
Oxygen and silicon atoms are represented by red and tan spheres, respectively.} \label{P1}
\end{figure}

\section{METHOD}

In the density functional theory (DFT) calculations of the structure of the SiO$_2$ nanowires we have
used two different simulation packages: the localized basis set code {\sc Fireball}~\cite{18}, used for the
full-characterization of the structures, and the plane-wave code {\sc Dacapo}~\cite{19,20}, used for refining
the electronic properties of the optimized structures. The computational {\sc Fireball} scheme, as well as
its theoretical foundations, has been described in full detail elsewhere~\cite{18,21,22}. {\sc Fireball} is
a supercell implementation of DFT formulated in real space using a basis of local pseudoatomic-orbitals.
Here we only summarize the main points. A self-consistent version of the Harris-Foulkes functional~\cite{23,24}
is used for solving the Kohn-Sham (KS) equations where the KS potential is calculated by approximating the
total charge by a superposition of spherical charges around each atom. The local density approximation (LDA)
is used for the description of exchange and correlation (XC) effects. An extended $sp^{3}d^{5}$ basis set of
single numerical atomic orbitals (one $s$, three $p$ and five $d$ orbitals) is employed for Si, and a double
numerical $sp^{3}s^{*}p^{*3}$ basis set for O. Core-electrons are replaced by norm-conserving
scalar relativistic pseudopotentials~\cite{25}.

\begin{figure}
\centerline{\includegraphics[width=12cm]{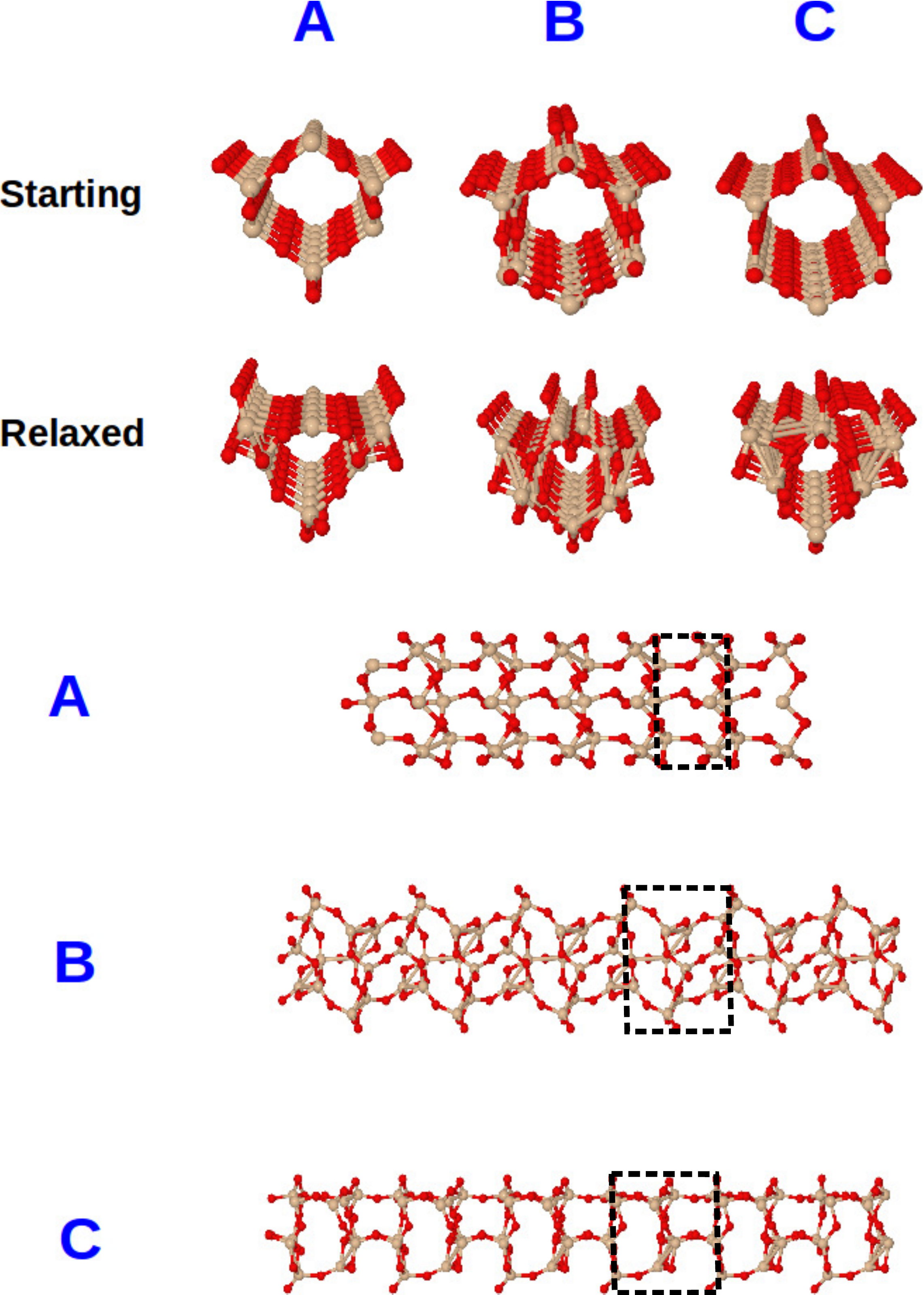}}
\smallskip \caption{(Color online) Front and lateral views of the most stable SiO$_2$ nanowires
obtained by relaxing fragments of bulk-phases (see~\ref{P1}): A) $\beta$-cristobalite;
B) ortho-tridymite; C) $\beta$-tridymite. Front views of the initial nanowires before relaxation
are also shown. Oxygen and silicon atoms are represented by red and tan spheres, respectively.
The black dashed boxes indicate the unit cell along the nanowire axis.}
\label{P2}
\end{figure}

In the {\sc Dacapo} code~\cite{19,20}, the KS equations are solved using a periodic supercell geometry. The
super-cells are such that the separation between neighbor nanowires is at least 15 \AA, in order to avoid
inter-nanowire interactions. The ion-electron interaction is modeled by ultra-soft pseudopotentials~\cite{26},
and XC effects are treated by the generalized gradient RPBE functional~\cite{19}. The one-electron valence
wave-functions are expanded in a basis of plane-waves, with energy cut-offs of 400 and 500 eV for the kinetic
energy and the electronic density, respectively. The supercell size and the energy cut-offs have been adjusted
to achieve sufficient accuracy in the total energy.

\begin{figure}
\centerline{\includegraphics[width=8cm]{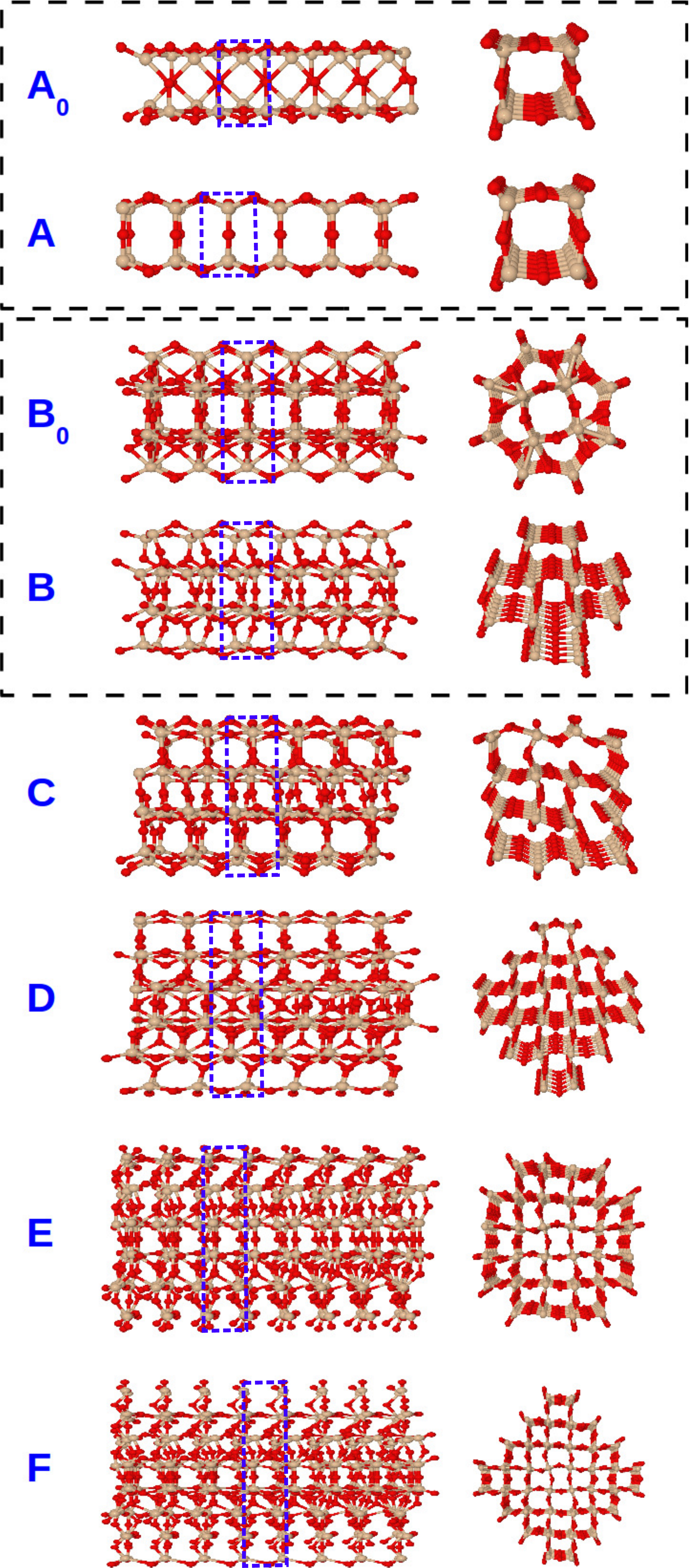}}
\smallskip \caption{(Color online) Lateral and front views of the most stable SiO$_2$ nanowires
with rutile-like structure. Views of the initial nanowires A and B before relaxation are also shown
(A$_0$ and B$_0$). From top to bottom (A to F) the nanowires have 4, 12, 16, 24, 32 and 40 SiO$_2$
functional units per unit cell, respectively. Oxygen and silicon atoms are represented by red and
tan spheres, respectively. The blue dashed boxes indicate the unit cell along the
nanowire axis.} \label{P3}
\end{figure}

The Brillouin zones of all the bulk-phases and nanowires were sampled with
6$\times$6$\times$6 and 12$\times$1$\times$1 Monkhorst-Pack grids~\cite{MPG}, respectively. In order to obtain the minimum energy structures
with {\sc Fireball}, a dynamical quenching technique has been employed~\cite{21,22}. Within this approach, all atoms are allowed to relax until an
energy minimum is reached. Then, the atoms are displaced slightly and the dynamical quenching process is continued. We run the dynamical relaxation process
until the force acting on each atom was less than 0.01 eV\AA$^{-1}$. At the same time, we minimized the strain on the unit cell in all
periodically repeated directions.

\begin{figure}
\centerline{\includegraphics[width=12cm]{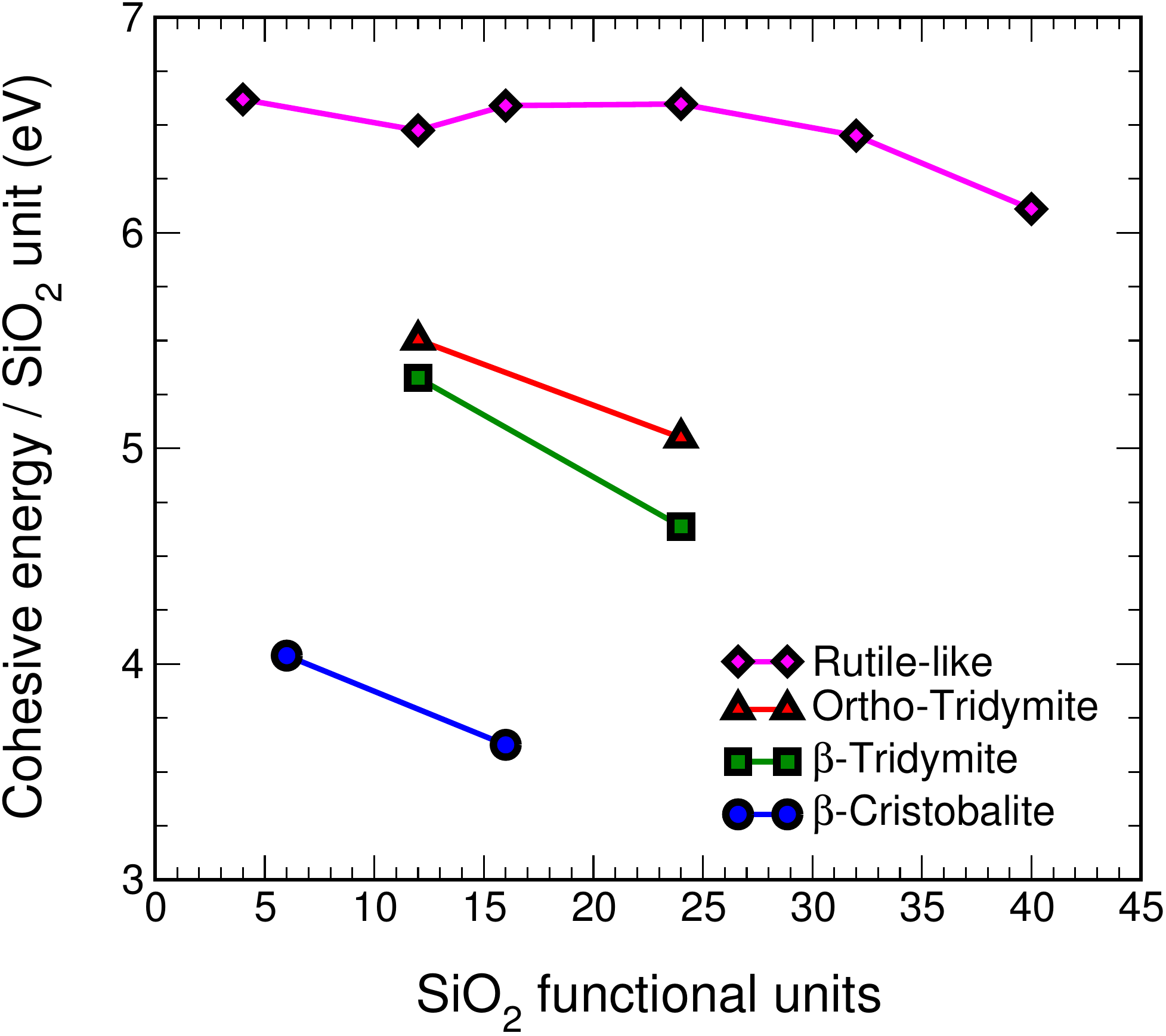}}
\smallskip \caption{(Color online) Cohesive energy of nanowires per SiO$_2$ functional
unit (as defined in the main text) as a function of the number of functional units in the unit cell,
for the nanowires shown in~\ref{P2} and~\ref{P3}.} \label{P4}
\end{figure}

Once the optimized structures have been established with the {\sc Fireball} code, these structures are used as starting geometries in a steepest-descent
re-optimization, carried out with the {\sc Dacapo} code. For all the systems, the convergence of the optimization procedure was immediate, revealing the
consistency between the two optimization approaches. Additionally, in order to check the influence of the temperature on the structure of the nanowires,
molecular dynamics simulations at temperature $T$=500 K have been performed on some of the most stable structures, which revealed the high structural
stability of the nanowires. In all cases, a typical dynamical pattern of two vibrational modes~\cite{9} is observed: a radial, breathing mode and a longitudinal,
stretching mode, conserving the total energy in time (in tests up to 10000 ps).

\section{GEOMETRICAL STRUCTURE OF THE NANOWIRES}

Due to the lack of information in the literature about the possible atomic structures of small diameter SiO$_2$ nanowires we have approached the problem from a
heuristic point of view. As a reasonable initial choice we constructed SiO$_2$ nanowires based on one dimensional-like fragments of the most stable bulk phases.
A number of polymorphs of silica, SiO$_2$, exist in nature. Coesite and stishovite are polymorphs which form at high pressure and temperature. At moderately lower
temperatures one finds the tridymite and cristobalite crystalline forms. These, like quartz, present low and high-temperature structural modifications. The crystalline
lattices of three of those phases, $\beta$-cristobalite~\cite{27}, ortho-tridymite~\cite{28}, and $\beta$-tridymite~\cite{29} are plotted in~\ref{P1}. According to this,
we have adopted as starting nanowire geometries fragments of the $\beta$-cristobalite, ortho-tridymite, and $\beta$-tridymite crystals in the form of SiO$_2$ nanowires.
Those fragments are of infinite length in the direction of the nanowire axis. The unit cell contains six SiO$_2$ units in the case of the $\beta$-cristobalite nanowire,
and twelve SiO$_2$ units in the two tridymite wires. Dynamical relaxations led to the optimized structures of~\ref{P2}, where frontal and lateral views of the most stable
nanowires are shown. \ref{P2} also shows a front view of the starting bulk-like nanowires before relaxation. Molecular dynamics simulations revealed that the minimum energy
structures obtained are robust and stable up to 500K at least. Optimization of the cell parameter in the axial direction was also allowed, but no significant variations
were observed with respect to the starting cell parameters. The calculated lattice parameters are 5.07 \AA, 8.27 \AA~and 8.24 \AA~for the $\beta$-cristobalite,
$\beta$-tridymite and ortho-tridymite-based SiO$_2$ nanowires, respectively. The starting bulk-like nanowires present a hexagonal-like section with some oxygen
atoms popping out due to stoichiometry. The section deforms after structural relaxation: the hexagonal inner cavity of the nanowire reduces its size and more oxygen
atoms appear on the outer part of the nanowire. The reconstruction is more severe in the two tridymite nanowires. We also did calculations in which the
wires were cut from the crystal choosing other directions for the wire axis. However, those led to less stable structures after optimization.

\begin{center}
\begin{table}
\begin{tabular}{l ccc ccc ccc}
\hline \hline
                          &&& $\hat{D}_{Si}$ (\AA) &&& $\hat{D}_{O}$ (\AA) &&& $\hat{d}_{Si-O}$ (\AA) \\ \hline \hline
\textbf{Bulk-based}       &&&                      &&&                     &&&                        \\ \hline
$\beta$-cristobalite      &&&  5.03                &&&  6.73               &&&  1.73$\pm$0.06         \\
Ortho-tridymite           &&&  6.31                &&&  8.74               &&&  1.74$\pm$0.08         \\
$\beta$-tridymite         &&&  6.43                &&&  8.45               &&&  1.74$\pm$0.08         \\ \hline
\textbf{Rutile-based}     &&&                      &&&                     &&&                        \\ \hline
4 SiO$_2$ units           &&&  4.67                &&&  6.57               &&&  1.73$\pm$0.02         \\
12 SiO$_2$ units          &&&  10.32               &&&  12.23              &&&  1.74$\pm$0.05         \\
16 SiO$_2$ units          &&&  14.12               &&&  15.64              &&&  1.74$\pm$0.04         \\
24 SiO$_2$ units          &&&  16.09               &&&  17.84              &&&  1.73$\pm$0.05         \\
32 SiO$_2$ units          &&&  18.67               &&&  21.16              &&&  1.74$\pm$0.06         \\
40 SiO$_2$ units          &&&  23.04               &&&  24.88              &&&  1.74$\pm$0.08         \\ \hline \hline
\end{tabular}
\caption{Average nanowire diameter estimated from the most external radially opposite Si atoms, $\hat{D}_{Si}$, average diameter estimated from the most
external radially opposite O atoms, $\hat{D}_{O}$, and average Si-O bond-length, $\hat{d}_{Si-O}$, in the bulk-based and
rutile-based SiO$_2$ nanowires. All distances are given in \AA. \label{T1}}
\end{table}
\end{center}

We have performed additional calculations starting with other structures. For this purpose we first took a narrow SiO$_2$ nanowire whose lateral and front views are shown in panel A$_0$ of~\ref{P3}. That structure, having four SiO$_2$ molecules per unit cell, is taken from previous works~\cite{8,9}
on nanowires of transition metal dioxides (MO$_2$), and it was obtained by relaxing nanowires cut from the rutile crystals. Shown in panel A is the
relaxed SiO$_2$ nanowire, whose structure is more symmetric than that of the starting nanowire. Next we relaxed a nanowire with the form of a two-wall
nanotube, whose lateral and front views are shown in panel B$_0$ of~\ref{P3}. This structure was also taken from previous work on transition metal
dioxide nanowires~\cite{8,9}. Structural relaxation distorts the original nanotube producing an atomic arrangement based on the rutile lattice (panel B).
Motivated by this feature, we explore a growth pattern to construct thicker nanowires based on the rutile-like structure of panel A. Nanowires with
16, 24, 32 and 40 SiO$_2$ units per cell are shown in panels C, D, E and F. These structures are shown after relaxation. For nanowires of a given size,
different starting configurations were considered, but only those leading to the most stable nanowires are shown in~\ref{P3}. The structures of all these
nanowires are stable in molecular dynamics simulations at 500K.

The cohesive energies of these nanowires per SiO$_2$ unit, defined as:
\be \label{CE}
E^{c}[\textrm{NW}] / SiO_2=\frac{E^{tot}[\textrm{NW}]-NE^{tot}[SiO_2 \textrm{~unit}]}{N},
\ee have been plotted versus the number of SiO$_2$ molecules per unit cell in~\ref{P4}. In ~\eqref{CE}, $E^{tot}[\textrm{NW}]$ is the total energy of the nanowire per unit cell, $N$ is the
number of SiO$_2$ molecules per unit cell, and $E^{tot}[SiO_2 \textrm{~unit}]$ is the total energy of an isolated gas-phase
SiO$_2$ molecule. The cohesive energies of the reconstructed ortho-tridymite and $\beta$-tridymite nanowires are similar, and higher than the cohesive energy
of the reconstructed $\beta$-cristobalite nanowires. That is, the tridymite nanowires are more stable.

In order to check the validity of the structures obtained in the simulations using the canonical (minimal) unit cell, we have carried out calculations taking a cell doubled in size, along the nanowire axis, compared to the original one. The aim of the test was to allow for the possibility of observing new structural reconstructions not seen in the dynamical quenches based on the minimum unit cell. The test was made for all the bulk-based nanowires and four rutile wires. The results obtained do not reveal any significant variation of the structures with respect to the nanowires obtained by just using the minimum unit cell.

The cohesive energies of the rutile-based nanowires are also included in~\ref{P4}. A similar value of the cohesive energy per SiO$_2$
unit is obtained for all the rutile-based nanowires, although it tends to decrease as the cross section of the nanowire increases. That is, the cohesive energy per SiO$_2$ unit tends
to decrease for thick nanowires. We think this is due to the fact that no stable bulk phases of SiO$_2$ with structures related to rutile exist at normal pressures.
The only bulk phase related to rutile is stishovite~\cite{30}, a high pressure phase with a mass density of 4.29 g cm$^{-3}$, much larger than the density of $\alpha$-quartz,
2.65 g cm$^{-3}$. Consequently, thick rutile-like nanowires are not expected to form at normal pressure. Thin nanowires, like those studied here, present fewer problems because the structural constraints
are weaker. Still, one can notice the enhanced stability of the rutile-like nanowires with respect to those obtained from the $\beta$-cristobalite, ortho-tridymite and
$\beta$-tridymite phases. The gain in cohesive energy with respect to the tridymite-like nanowires with a similar number of SiO$_2$ units ranges between 1 eV and 2 eV per SiO$_2$ unit, and from 2.5 and 3 eV per SiO$_2$
unit with respect to the $\beta$-cristobalite-like nanowire.

To interpret the behaviour of the cohesive energies in~\ref{P4} one has to notice that the range of cross sections for these nanowires corresponds to the non-scalable regime, well known in clusters~\cite{Alonso}, where the properties vary in a non-smooth way, often showing oscillations, as the cluster size increases. The present calculations suggest that the non-scalable regime occurs also in thin nanowires, and is responsible for the oscillations of the cohesive energy of the rutile-like wires, and for the initial decrease of the cohesive energies of the $\beta$-cristobalite, ortho-tridymite and
$\beta$-tridymite wires.

Basic structural measurements of a nanowire can include its diameter as well as its length. These parameters may be related to laboratory preparation
of nanowires~\cite{13,14,15}. At the atomistic level, a characteristic parameter would be an average bond-length. Accordingly,~\ref{T1} provides the calculated average diameter
estimated from the most external radially opposite Si atoms, $\hat{D}_{Si}$, the average diameter estimated from the most external radially opposite O atoms, $\hat{D}_{O}$, and the average Si-O bond-distance,
$\hat{d}_{Si-O}$, in all the bulk-based and rutile-based SiO$_2$ nanowires. Notice that the largest diameter obtained is that for $\hat{D}_{O}$ in the case of
40 functional units per unit cell. Its value is about 25 \AA=2.5 nm, which, from the point of view of experimental synthesis, corresponds to a thin nanowire. Although SiO$_2$ nanowires
used as a core in composite systems such as RuO$_2$/SiO$_2$ have tipically been one order of magnitude thicker~\cite{13}, there is no reason
why smaller core nanowires could not be processed and employed. Since the thicknesses of the nanowires studied here are on the non-scalable regime, where the properties may vary in a
non smooth way, it is not easy to extrapolate the results of~\ref{P4} to very thick wires.

\section{ELECTRONIC STRUCTURE}

\subsection{Density of states (DOS)}

\begin{figure}
\centerline{\includegraphics[width=12cm]{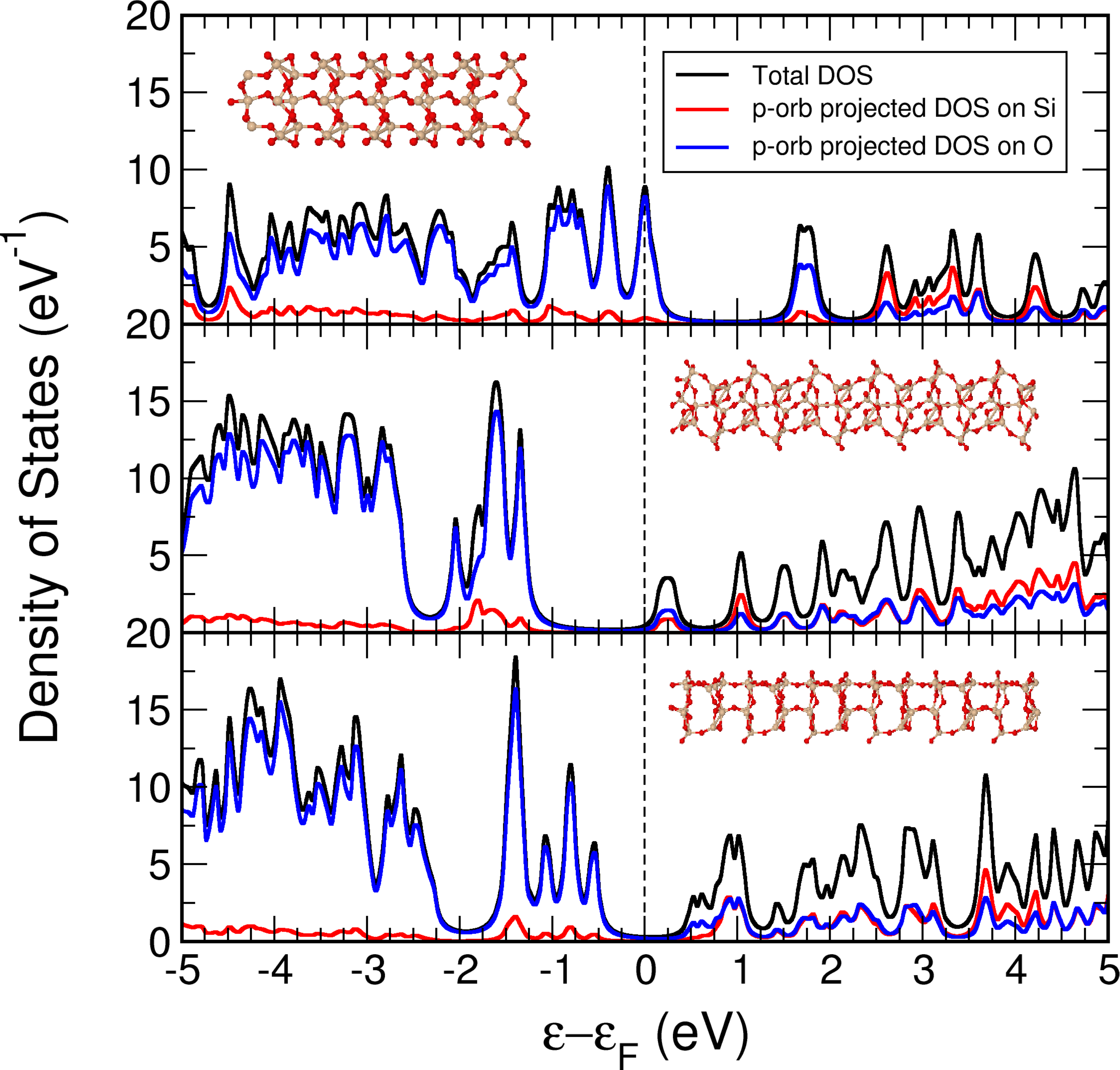}}
\smallskip \caption{(Color online) Density of electronic states (DOS) versus energy for the SiO$_2$
nanowires of~\ref{P2}. The Fermi energy $\varepsilon_F$ is taken as reference. Upper, middle
and lower panels correspond to nanowires based on the $\beta$-cristobalite, ortho-tridymite and
$\beta$-tridymite crystals, respectively. Black curves give the total DOS, and red and blue
curves represent DOS projected, respectively, onto the $p$ states of Si and O atoms. Lateral views
of the nanowires are included as insets. Oxygen and silicon atoms are represented by red and tan
spheres, respectively.} \label{P5}
\end{figure}

\begin{figure}
\centerline{\includegraphics[width=12cm]{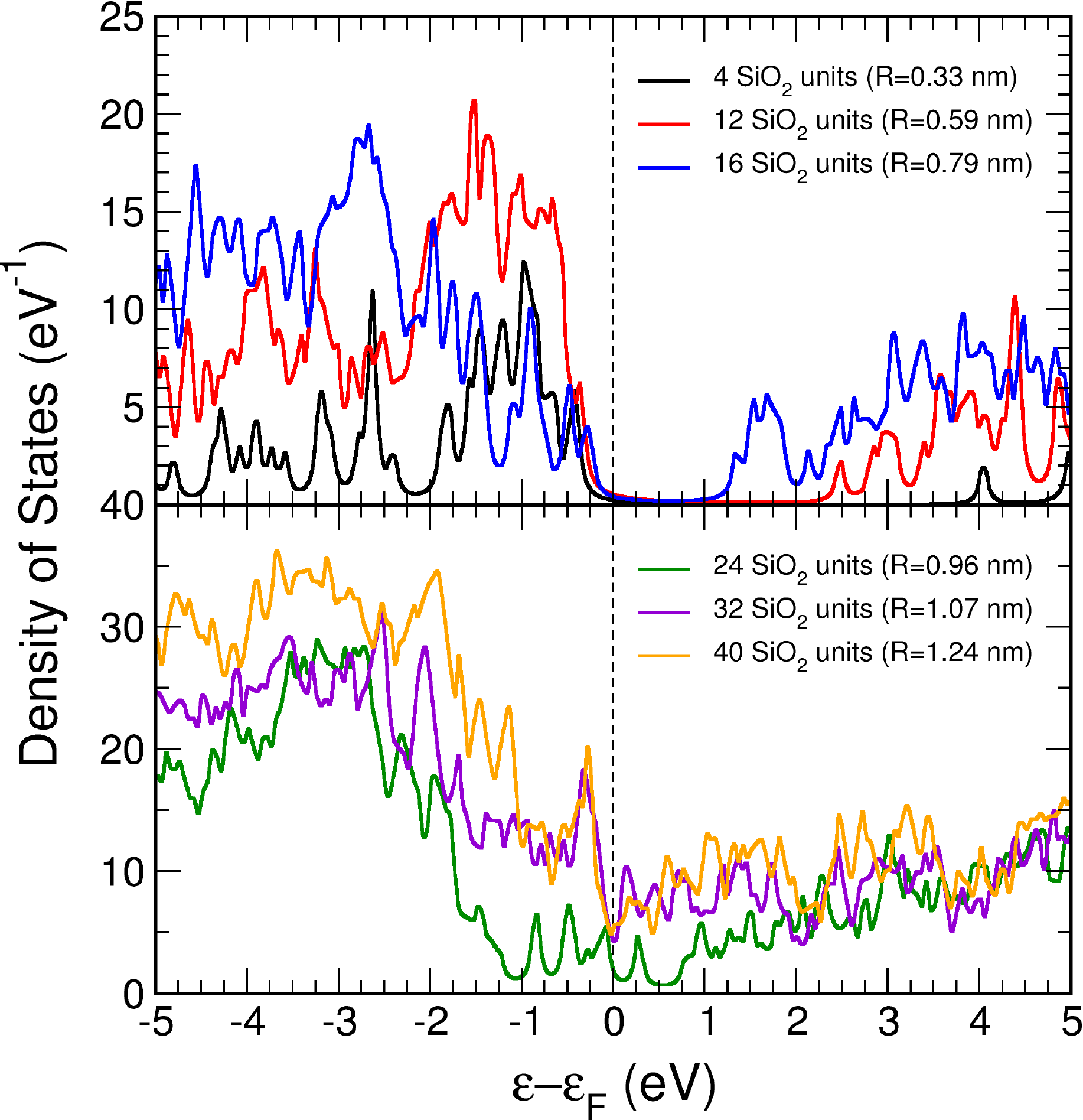}}
\smallskip \caption{(Color online) Density of electronic states (DOS) versus energy for the
rutile-based SiO$_2$ nanowires of~\ref{P3}. The Fermi energy $\varepsilon_F$ is taken as
reference. The number of SiO$_2$ units per cell and the average radius of each nanowire are
indicated in the inset.} \label{P6}
\end{figure}

\begin{figure}
\centerline{\includegraphics[width=15cm]{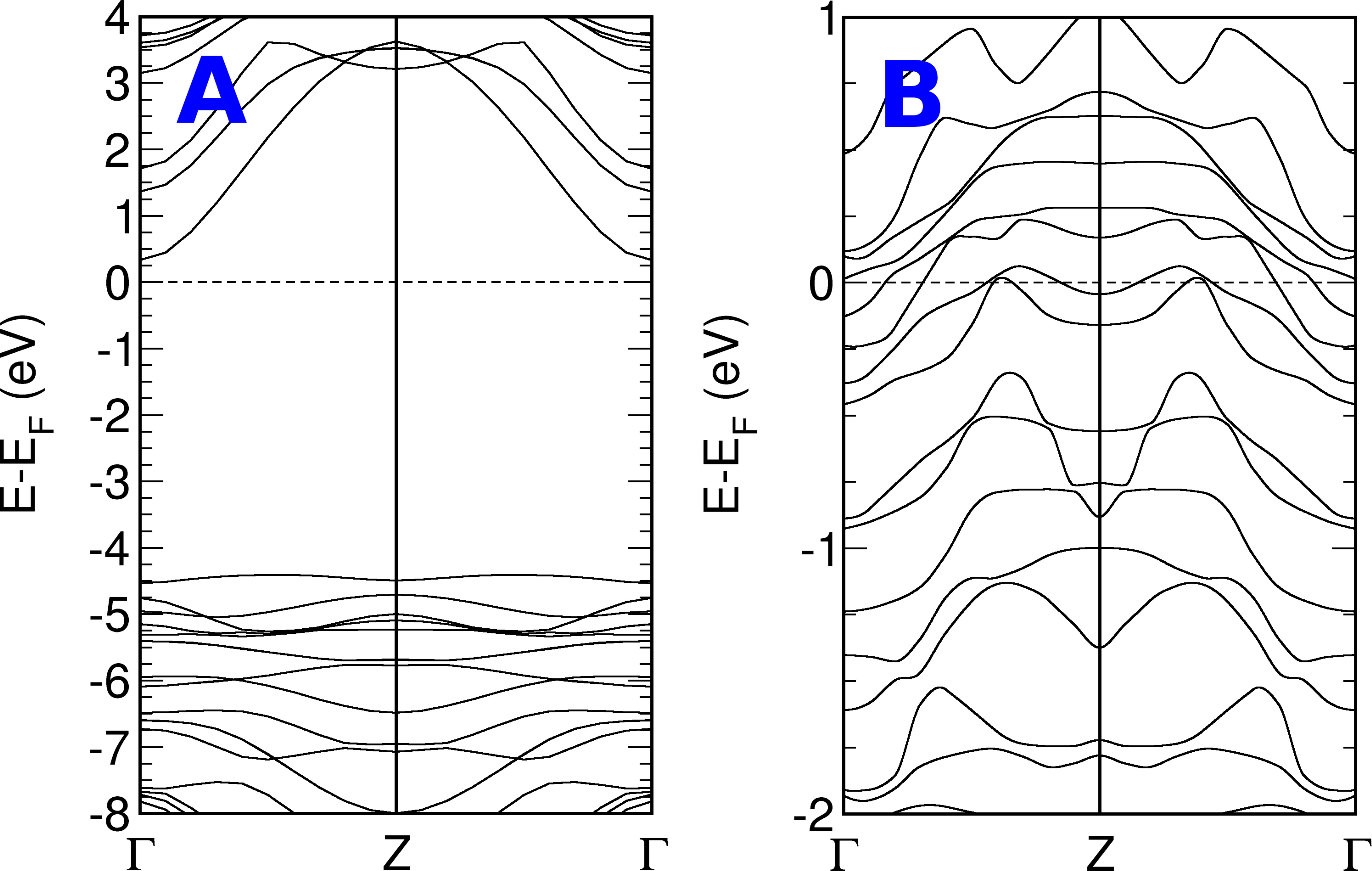}}
\smallskip \caption{(Color online) Band-structure diagrams (referred to the Fermi energy level) for
rutile-based SiO$_2$ nanowires (see~\ref{P3}) with 4 (panel A) and 24 (panel B) functional units
along the most relevant symmetry line from $\Gamma \rightarrow$ Z (along nanowire axis).} \label{P6a}
\end{figure}

We begin our analysis by considering the nanowires obtained from the cristobalite and tridymite bulk-phases (see~\ref{P2}). In addition to its intrinsic interest,
the study provides insight on how structural changes in the nanowires may affect their electronic properties, bonding character and charge transfer between atoms.
Such information could prove useful when trying to form SiO$_2$ semiconducting nanostructured materials, as well as when forming composite nanocables with overlying
metal or semiconducting materials.

\ref{P5} shows the total electronic DOS profiles versus energy for the three nanowires of~\ref{P2}. In order to mimic the experimental resolution~\cite{30a}, a Lorentzian broadening
of 0.05 eV has been applied to the energy levels. The magnitude of the electronic gaps, measured as the difference between the energies of the highest occupied (HOMO) and
lowest unoccupied (LUMO) molecular orbitals, is 1.69, 1.57 and 1.06 eV for the $\beta$-cristobalite (top panel), ortho-tridymite (middle panel) and $\beta$-tridymite
(bottom panel) nanowires, respectively, revealing the potential semiconducting-like character of these nanostructures, especially if appropriate donor or acceptor atoms
were available to reside in shallow states below the band edges within the bandgap. The calculated band-gaps are smaller than the gaps of the corresponding bulk crystals.
The gap of bulk $\beta$-cristobalite obtained by DFT is in the range 5.5-5.7 eV for LDA calculations, and increases to 10.1-10.3 eV if quasiparticle corrections are taken
into account~\cite{31}. A gap of 6.6 eV is obtained in gradient-corrected BLYP calculations~\cite{32}. Our RPBE calculations give a gap of 7 eV. Turning to the tridymite
crystals, the BLYP gap of $\beta$-tridymite is 6.6 eV~\cite{32}, and our RPBE gaps for ortho- and $\beta$-tridymite are 5.8 and 7.7 eV, respectively. That is, the bulk
phases are insulators. A substantial lowering of the gap of the nanowires with respect to the bulk crystals is then obtained. The band-gap
values of the nanowires are similar to those for some elemental bulk semiconductors like Si, or III-V binary and ternary semiconductor compounds, like GaAs, InGaP, and InGaAs,
with varying stoichiometry~\cite{33,34,35}.

\begin{figure}
\centerline{\includegraphics[width=15cm]{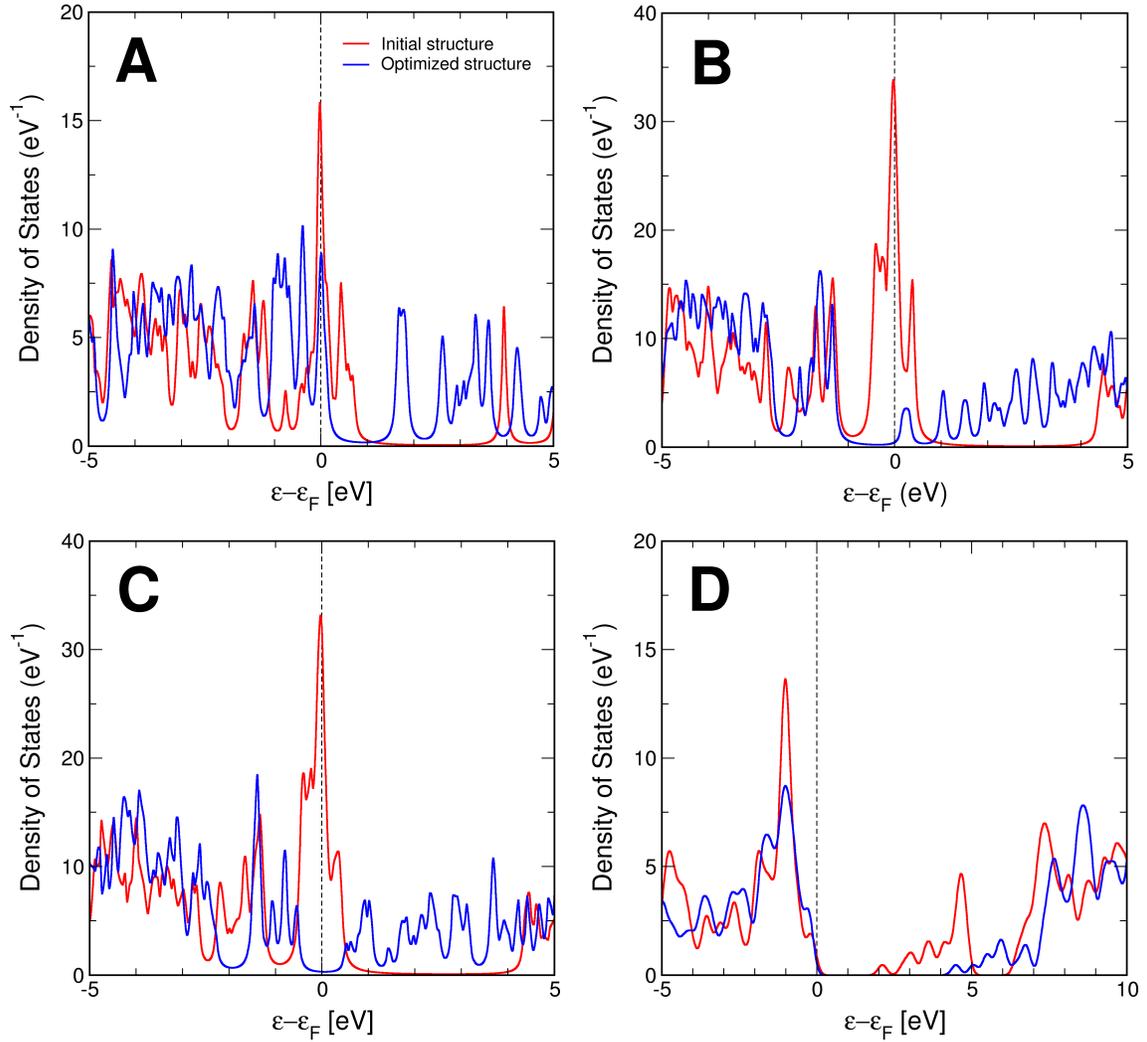}}
\smallskip \caption{(Color online) Density of electronic states (DOS) versus energy
for the $\beta$-cristobalite (A), ortho-tridymite (B) $\beta$-tridymite (C) (see~\ref{P2}), and the
smallest rutile-based (D) (see~\ref{P3}) SiO$_2$ nanowires. The Fermi energy $\varepsilon_F$ is
taken as reference. In each panel, DOS profiles are shown for the structures before and after structural
relaxation.} \label{P6b}
\end{figure}

Regarding the DOS profiles of the three nanowires we observe different band structure behaviors: i) the $\beta$-cristobalite nanowire (top panel) acts as a pure acceptor
semiconductor-like material, able to give up mobile holes whose volumetric density is
\be \label{eq1}
p=U_{v}e^{-(\varepsilon_F-\varepsilon_v)/k_{B}T}\approx U_v,
\ee since $\varepsilon_v\approx\varepsilon_F$ (here $\varepsilon_v$ is the top of the valence band, $k_B$ is the Boltzmann constant and $U_v$ is a numerical constant
depending on the temperature $T$); ii) the ortho-tridymite nanowire (middle panel) acts as a potential donor semiconductor-like material, able to create mobile electrons
with density
\be \label{eq2}
n=U_{c}e^{-(\varepsilon_c-\varepsilon_F)/k_{B}T}\approx U_c,
\ee since $\varepsilon_c\approx\varepsilon_F$ (here $\varepsilon_c$ is the bottom of the conduction band and $U_c$ is a numerical constant depending on $T$); and iii) the
$\beta$-tridymite nanowire (bottom panel) acts as an intrinsic semiconductor with tiny carrier densities
\be \label{eq3}
n_0=p_0=\sqrt{U_vU_c}~e^{-(\varepsilon_c-\varepsilon_v)/2k_{B}T},
\ee with the Fermi level located in the middle of the gap
\be \label{eq4}
\varepsilon_F=\frac{\varepsilon_v+\varepsilon_c+k_{B}T\ln{(U_v/U_c)}}{2}\approx\frac{\varepsilon_v+\varepsilon_c}{2}
\ee for $T\rightarrow$ 0 or $U_v/U_c\approx$1, and with no preference to donate or accept electronic charge~\cite{36}.

\begin{figure}
\centerline{\includegraphics[width=12cm]{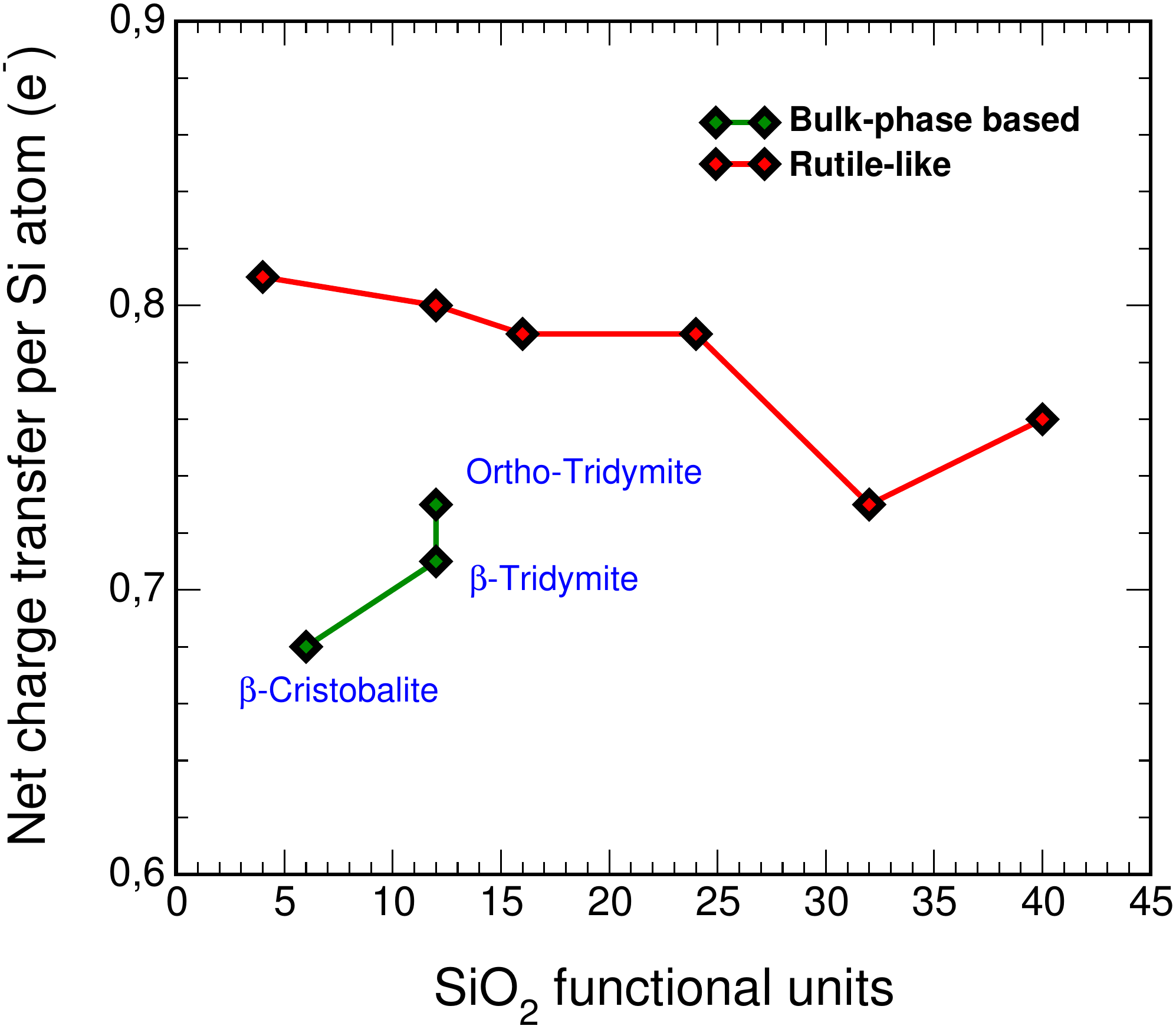}}
\smallskip \caption{(Color online) Average charge transfer per Si atom, in e$^{-}$, from Si to
O atoms, in the SiO$_2$ nanowires of~\ref{P2} and~\ref{P3}.} \label{P7}
\end{figure}

In~\ref{P5} we also show the projected DOS onto the $p$-states of Si and O atoms. In the energy range plotted, the occupied part of the DOS is dominated by the $p$-states of O.
This means that the oxygen atoms have accepted some electronic charge from the silicon atoms (details are provided below). The sum of the Si and O $p$-orbital contributions
falls slightly below the total DOS, the difference being the smaller $s$-orbital contributions.

\begin{center}
\begin{table}
\begin{tabular}{l c c c}
\hline \hline
\textbf{Transport gaps (eV)}            & Standard DFT         & quasi-particle G$_0$W$_0$ & GW + BSE \\ \hline \hline
\textit{Rutile-like nanowire}           &                      &                           &          \\ \hline
4 SiO$_2$ units                         &  4.53                &  4.80                     &  5.98    \\
12 SiO$_2$ units                        &  2.84                &  2.98                     &  3.58    \\
16 SiO$_2$ units                        &  1.62                &  1.73                     &  2.01    \\ \hline \hline
\end{tabular}
\caption{Transport gaps (in eV), calculated within standard DFT, quasi-particle G$_0$W$_0$ and
GW + BSE formalisms, for the rutile-like nanowires with 4, 12 and 16 SiO$_2$ functional units.} \label{T1b}
\end{table}
\end{center}

\ref{P6} shows the total DOS profiles versus energy for the most stable rutile-based nanowires of~\ref{P3}. As observed in the top panel of~\ref{P6}, thin rutile-like nanowires (with 4, 12 and 16  SiO$_2$  molecules per unit cell) show a semiconducting behavior, and the electronic gap is progressively reduced as the number of SiO$_2$ functional
units in the cell increases. For thicker nanowires, those in the bottom panel, new peaks appear in the DOS near the Fermi energy, and this region becomes congested. The structure of the electronic levels in that region can be seen in~\ref{P6a}. That figure shows the band structure of two nanowires, with 4 and 24 SiO$_2$ functional
units in the cell, respectively. The left-side panel shows clearly the gap existing in the thin nanowire. On the other hand, the band structure of the thick nanowire (notice the different energy scale) shows that a bunch of states occupy the region of the gap. These can be interpreted as surface-like states.

A comparison between~\ref{P5} and~\ref{P6} shows differences between the electronic gaps corresponding to nanowires of similar thickness but different structures.
The $\beta$-cristobalite nanowire (with four SiO$_2$ units per cell) has a gap of 1.69 eV, while the gap of the thinnest rutile-like nanowire (six SiO$_2$ units per cell)
is 4.93 eV. For nanowires with 12 SiO$_2$ units per cell, the differences are again evident, although not so large: the rutile-like nanowire has a gap of 2.91 eV, and the
two tridymite nanowires show gaps of 1.57 and 1.06 eV. The conclusion is that by varying the thickness and the structure of the nanowires, these can be engineered to produce
a desired value of the electronic gap.

There is a reduction of the band-gap of the nanowires with respect to the SiO$_2$ bulk crystals. We ascribe the difference to the particular structure
of the nanowires, with many unsaturated oxygen atoms on the surface. This is required by stoichiometry, two O atoms per Si atom, and it becomes reflected, as shown in~\ref{P5},
in the dominance of the occupied DOS near the Fermi level by the $p$-like states of oxygen. The reduced gaps also indicate that the nanowires will be more reactive than the
usual bulk SiO$_2$ phases.

The structural relaxation of the nanowires shown in~\ref{P2} and ~\ref{P3} is well explained by looking at the DOS. ~\ref{P6b} shows the DOS profiles of four nanowires in the original unrelaxed structure and after structural optimization. In the cases of  $\beta$-cristobalite (panel A), ortho-tridymite (panel B) and $\beta$-tridymite (panel C), the density of states in the unrelaxed structure shows several high peaks in the region near the Fermi energy. Those peaks correspond to electronic states located inside a gap of width about 5 eV, which reminds the gap in the bulk crystals. The high DOS near the Fermi energy indicates that the unrelaxed structures of the nanowires are unstable. In fact, as the initial structures of~\ref{P2} relax to more stable arrangements, the states at the Fermi level almost disappear, and a true gap develops, although smaller in magnitude than the original gap of the crystal. At difference with the previous wires, the smallest rutile-based nanowire shows a gap and no peaks at the
Fermi energy. The effect of structural relaxation is, in this case, to widen the gap by moving the unoccupied electronic states up in energy. The widening of the gap increases the stability of the nanowire. However, as discussed above, surface states appear in the gap for thick rutile-like nanowires.

In order to check the reliability of the DFT-DOS profiles and transport gaps, we have carried out more accurate 
electronic structure calculations for the rutile-based nanowires with 4, 12 and 16 functional units. The choice of these testing 
structures is based on: i) their transport gaps are easy to identify (see~\ref{P6}), while the thicker nanowires show states 
inside the gap region; ii) their small cell sizes lower the substantial computational effort of such calculations. For this 
purpose, we have calculated the many-body corrections using two different formalisms: a) quasi-particle corrections have been 
calculated within the G$_0$W$_0$ formalism~\cite{37,38} (quasi-particle Green function screened interaction approach) to properly 
account for the exchange-correlation self-energy; b) the GW + Bethe-Salpeter equation (BSE) approximation~\cite{38a,38b,38c}, 
where the quasi-particle GW approach is further corrected to account for excitonic effects (electron-hole interation) in the 
unoccupied and excited states. The GW + BSE formalism includes, by construction, excitonic effects, which are important in highly 
correlated systems, and combines the excellent performance of the GW + standard DFT approach for the ground state electronic 
structure with the high accuracy of the BSE methodology for the calculation of excitations. These calculations have been carried 
out with the {\sc Yambo} package~\cite{yambo}. More than 200 unoccupied electronic bands were necessary to obtain converged 
self-consistent results.

\ref{T1b} shows the transport gaps (in eV) predicted by the G$_0$W$_0$ and GW + BSE formalisms for the three 
rutile-like nanowires. The comparison between the DFT and G$_0$W$_0$ gaps indicates that the first-order quasi-particle corrections 
are small: the G$_0$W$_0$ gaps are larger than the DFT gaps by 0.27, 0.14 and 0.11 eV for the nanowires with 4, 12 and 16 SiO$_2$ 
units, respectively. The excitonic corrections are larger. These lead to an additional increase of the gap by 1.18, 0.60 and 0.28 
eV for the same three nanowires, respectively. The percentage change of the gap due to self-energy corrections, 
\be \label{change1}
\Delta E^{DFT\rightarrow G_0W_0}_g(\%)=100\times\frac{E^{G_0W_0}_g - E^{DFT}_g}{E^{DFT}_g},
\ee is nearly independent of wire diameter (about 6 \%). On the other hand, the additional percentage 
change induced by excitonic effects, 
\be \label{change2}
\Delta E^{G_0W_0\rightarrow GW + BSE}_g(\%)=100\times\frac{E^{GW + BSE}_g - E^{G_0W_0}_g}{E^{G_0W_0}_g},
\ee decreases as the nanowire diameter increases: 25 \%, 20 \% and 16 \%, for the three nanowires, respectively. 
Thus, the largest gap corrections due to many-body effects occur for the thinnest nanowires. It is worth noticing, however, 
that the trend of a decreasing gap with increasing nanowire diameter predicted by the standard DFT calculations is maintained 
after including many-body corrections.

\subsection{Electron distribution and charge transfer}

\begin{center}
\begin{table}
\begin{tabular}{l c c c}
\hline \hline
                           & $\beta$-cristobalite & Ortho-tridymite & Rutile \\ \hline \hline
\textbf{Si atom}           &                      &                 &        \\ \hline
$s$ orbital charge         &  1.1                 &  1.0            &  1.0   \\
$p$ orbital charge         &  2.2                 &  2.2            &  2.2   \\
Net charge gain / lost     & -0.7                 & -0.8            & -0.8   \\ \hline
\textbf{O atoms}           &                      &                 &        \\ \hline
$s$ orbital charge         &  3.4                 &  3.4            &  3.3   \\
$p$ orbital charge         &  9.3                 &  9.4            &  9.5   \\
Net charge gain / lost     &  0.7                 &  0.8            &  0.8  \\ \hline \hline
\end{tabular}
\caption{Average $s$ and $p$ orbital charges (in e$^{-}$ units) in the Si and O atoms, and net
electronic charge gained / lost (in e$^{-}$ units), for the $\beta$-cristobalite, ortho-tridymite
and the rutile-like nanowire with 24 functional units. Notice that the values shown correspond to
one Si atom and two O atoms, that is, to a SiO$_2$ unit. \label{T2}}
\end{table}
\end{center}

In order to understand the bonding mechanism we have carried out an analysis of the electronic charges of the atoms. Theoretical calculations giving the electronic
charge distribution in the system can, in principle, provide this type of information, but the way to extract it is not unique. The usual output of quantum mechanical
calculations is a continuous electronic charge density and it is not evident how to partition electrons amongst fragments of the system such as atoms or molecules.
Different schemes have been proposed, some of them based on localized electronic orbitals~\cite{39,40} and others based on the electron density~\cite{41}.

Taking advantage of the localized basis set used in the {\sc Fireball} code, we can evaluate the electronic charge in each localized orbital at the end of the
self-consistent calculation~\cite{18,21,22}. The L\"owdin method~\cite{39,40} has been employed to obtain the orbital charges. With those values, the sign and
the magnitude of the net charge transfer between atoms can be obtained. In order to check the accuracy of the results we have compared the atomic charges with
the Bader atomic charges~\cite{41}, obtained using electron densities from the plane-wave {\sc Dacapo} code. There is excellent agreement between atomic charges
calculated in the two ways.

\begin{figure}
\centerline{\includegraphics[width=12cm]{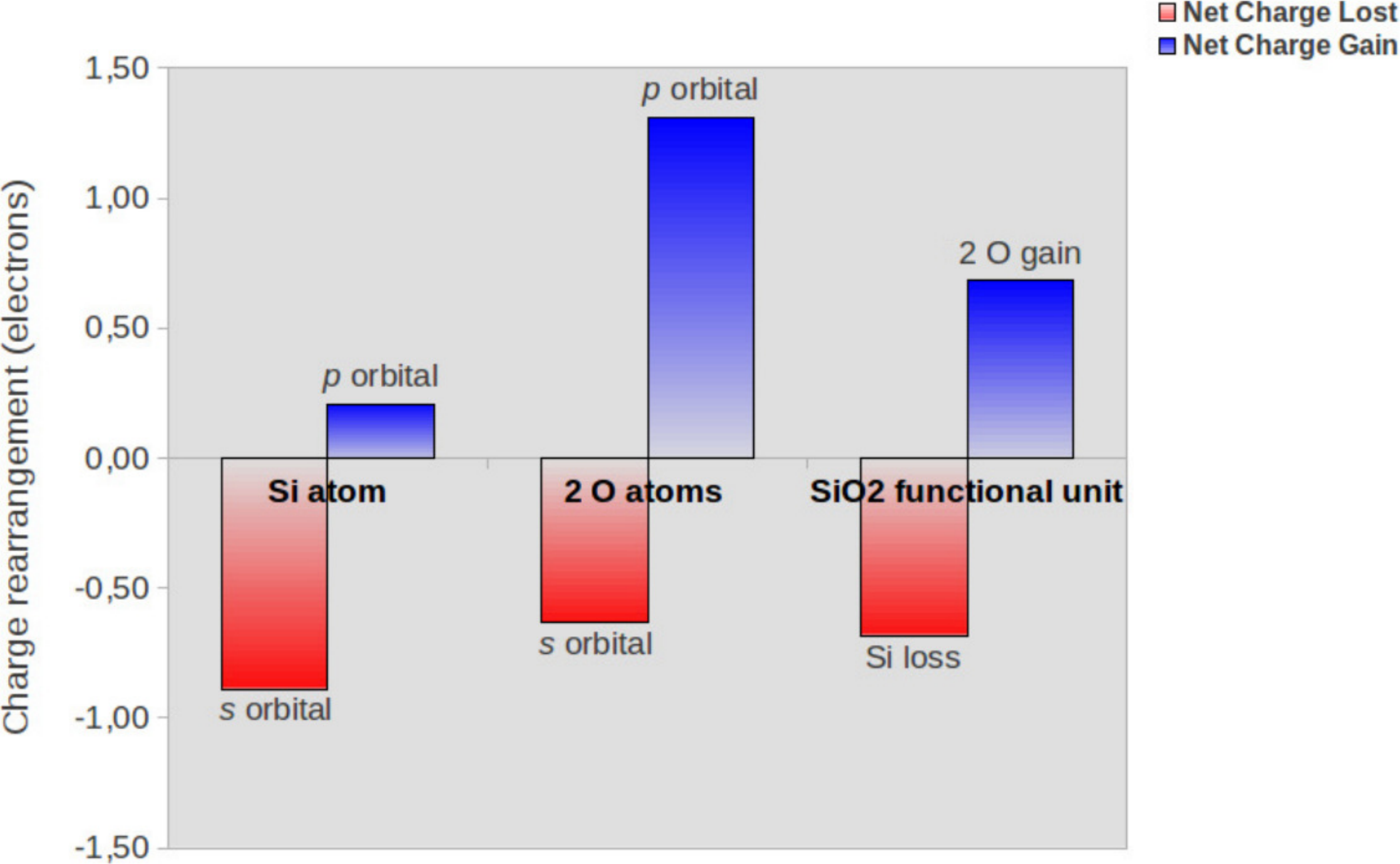}}
\smallskip \caption{(Color online) Bars diagram showing the $s$ and $p$ orbital electronic
rearrangements in the Si and O atoms of the $\beta$-cristobalite-like nanowire (in e$^{-}$ units),
per SiO$_2$ functional unit. The net charge transfer is also shown. Blue and red filling indicates
electronic charge lost and gained, respectively.} \label{P8}
\end{figure}

Electronic charge is transferred from Si to O atoms due to the higher electronegativity of oxygen. \ref{P7} shows the average net charge transfer, in e$^{-}$ units,
from Si to O atoms, per SiO$_2$ unit. In other words, this is the average charge transfer per Si atom. Charge transfers in the rutile-like nanowires range between
0.73 and 0.81 e$^{-}$ per Si atom. These are slightly larger than the charge transfers in the cristobalite- and tridymite-like nanowires, 0.68-0.73 e$^{-}$. The
larger charge transfer in the rutile-like nanowires correlates with their higher stability. For this purpose, one can notice the qualitative similarity of the data in~\ref{P4} and~\ref{P7}.

\ref{T2} shows average $s$ and $p$ orbital charges, in e$^{-}$ units, in the Si and O atoms for three nanowires: $\beta$-cristobalite and ortho-tridymite wires, and
the rutile-like nanowire with 24 SiO$_2$ units per cell. The net electronic charges lost (by each Si atom) or gained (by two O atoms) are also given. The results for
the two tridymite-like nanowires are very similar. Also the results for the six rutile-like nanowires are very similar. This, plus the information in~\ref{T2},
indicates that the average atomic charge populations, total and orbital-resolved, are similar in all the nanowires. We then conclude that the chemical bonding in
these nanowires exhibits common features, quite independent of the detailed structure of the nanowires.

The chemical bonding can be described as a mixture of covalent and ionic contributions. In all cases, the $p$-like electron population of the O atoms
gains 0.70-0.75 e$^{-}$ per O atom (1.4-1.5 e$^{-}$ per SiO$_2$ unit). The Si atoms also increase their $p$-like charge by 0.1-0.2 e$^{-}$ per atom. This general
increase of $p$-like population in the nanowires arises from a loss of charge in the $s$-like orbitals (0.9 e$^{-}$ lost by the Si atoms and 0.30-0.35 e$^{-}$ lost
by each O atom). This is summarized in the bar diagram of~\ref{P8} for the $\beta$-cristobalite nanowire.

\begin{figure}
\centerline{\includegraphics[width=8cm]{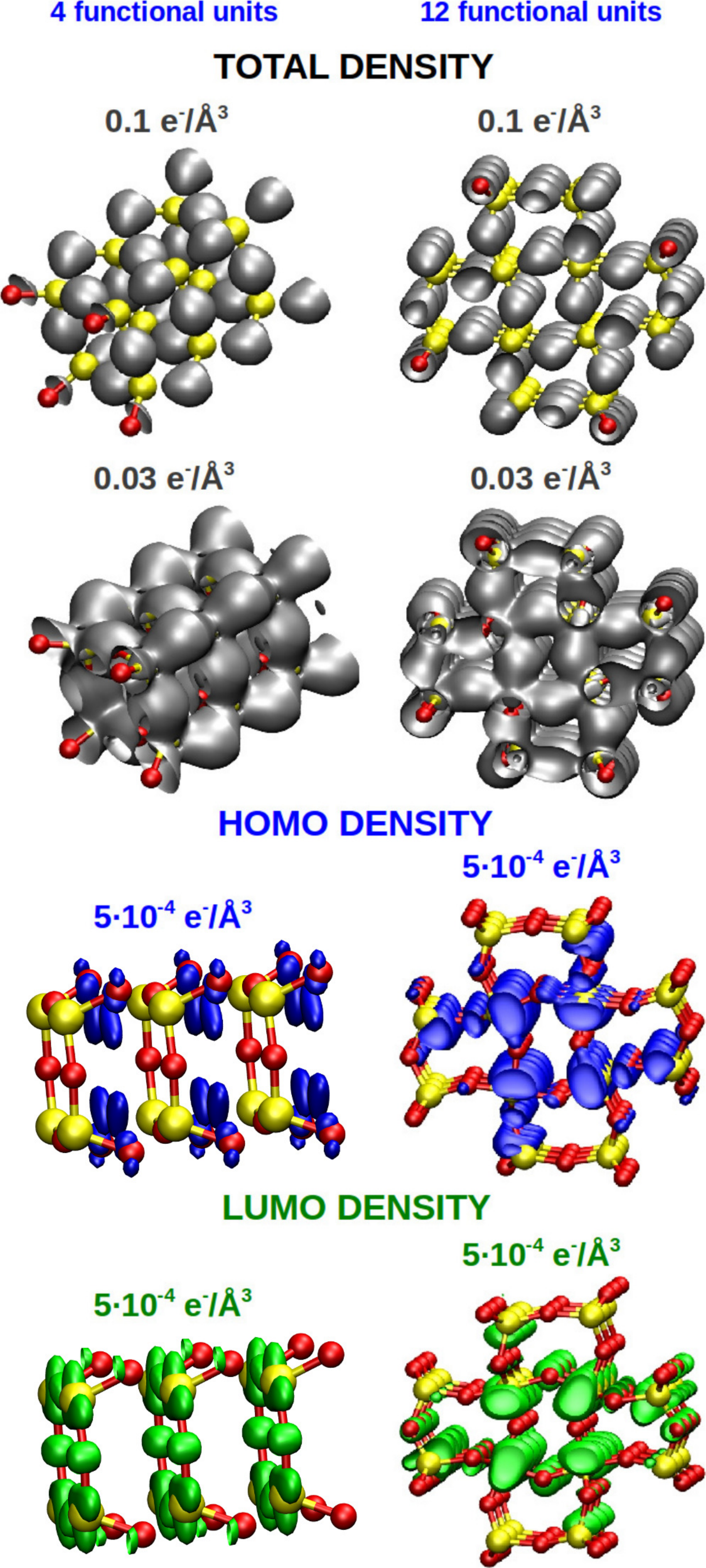}}
\smallskip \caption{(Color online) Surfaces of equal electron density for two rutile-like nanowires
with four and twelve SiO$_2$ units per unit cell. The densities selected have the values 0.1 and
0.03 e$^{-}\textrm{\AA}^{-3}$ for the total densities, and 5$\cdot$10$^{-4}$ e$^{-}\textrm{\AA}^{-3}$
for the HOMO and LUMO densities. Oxygen and silicon atoms are represented by red and yellow spheres,
respectively, and the iso-density surfaces by grey, blue and green shiny tones for the total, HOMO
and LUMO densities, respectively.} \label{P9}
\end{figure}

In order to visualize the spatial distribution of the electronic charge,~\ref{P9} shows surfaces of constant electron density in 3 dimensional perspective for the
rutile-like nanowires with 4 and 12 SiO$_2$ units per cell. The densities selected are 0.1, and 0.03 e$^{-}\textrm{\AA}^{-3}$. The results are similar for the two
nanowires (and for others not shown here). The surfaces corresponding to a density of 0.1 e$^{-}\textrm{\AA}^{-3}$ are formed by a set of disconnected pieces showing
a concentration of charge around the oxygen atoms, aided by the charge transfer from Si to O atoms. The surface corresponding to the lower density, 0.03
e$^{-}\textrm{\AA}^{-3}$, is a connected surface displaying the mixture of ionic and covalent bonding in the nanowires. \ref{P9} also shows the charge densities
of the HOMO and LUMO orbitals, again for the rutile-like nanowires with 4 and 12 SiO$_2$ units per cell (all with values of 5$\cdot$10$^{-4}$ e$^{-}\textrm{\AA}^{-3}$).
In the thin nanowire, the HOMO orbital is localized near the planes formed by the most external oxygen atoms (each of those planes contains four oxygen atoms).
In contrast, the LUMO orbital is localized near the planes containing the distorted squares formed by four Si and four O atoms. For the thicker nanowire,
the HOMO orbital is placed in between the planes containing the Si atoms, and the LUMO orbital mostly surrounds internal Si atoms.

\section{SUMMARY}

This work presents ab-initio simulations for SiO$_2$ nanowires. The structures of the nanowires have been determined: i) by relaxing nanowires cut from SiO$_2$ bulk
crystals ($\beta$-cristobalite, ortho-tridymite and $\beta$-tridymite), and ii) by relaxing nanowires with a rutile-like structure. The analysis of the electronic
states indicates that many of these nanowires have a semiconducting character, different from the insulator character of the macroscopic bulk-crystals. However, for the thick rutile-like nanowires we found the gap congested by the presence of surface states. The bonding in all the nanowires can
be described as partly ionic and partly covalent. Novel uses for nanoscopic SiO$_2$ nanowires may arise from insights gained about the electronic density of states,
chemical bonding, charge density distribution and cohesive energies per SiO$_2$ unit as a function of nanowire thickness and structure.

\section{ACKNOWLEDGEMENTS}

Results presented here were conducted under the aegis of the NRL NanoScience Institute (NSI), project
"Nanoparticulate Ruthenium Dioxide Shells on Dielectric Cores: Basic ElectroChemistry, Physics, and Material
Science of a Single-Unit-Thick Electron Conductor and the Implications for Energy and Electro-Optical Applications".
The work was supported by Spanish MICINN and the European
Regional Development Fund (grants MAT2008-06483-C03-01, MAT2011-22781 and FIS2010-16046),
Junta de Castilla y Le\'on (grant VA158A11-2), CAM (grant S2009/MAT-1467),  and the European Project MINOTOR (Grant FP7-NMP-228424).
J.I.M. acknowledges funding from Spanish MICIIN through Juan de la Cierva Program.



\newpage

\begin{figure}
\centerline{\includegraphics[width=\columnwidth]{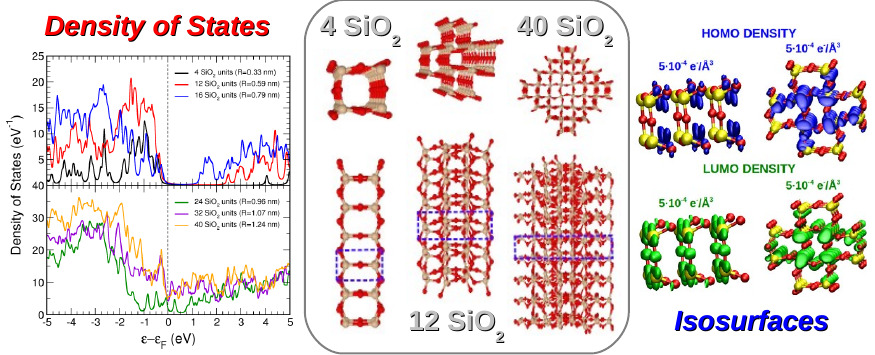}}
\smallskip \caption{(Color online) Table of contents.} \label{PTOC}
\end{figure}


\begin{thebibliography}{50}


\bibitem{1} Heller, A. \emph{Science} {\bf 1984}, \textit{223}, 1141-1148.
\bibitem{2} Murphy, J.; Srinivasan, S.; Conway, B. E. \emph{Electrochemistry in Transition: From the 20$^{th}$ to the 21$^{st}$ Century}; Plenum: New York, 1992.
\bibitem{3} Khaselev, O.; Turner, J. A. \emph{Science} {\bf 1998} \textit{280}, 425-427.
\bibitem{4} Gr\"atzel, M. \emph{Nature} {\bf 2001} \textit{414}, 338-344.
\bibitem{5} Whitesides, G. M.; Crabtree, G. W. \emph{Science} {\bf 2007} \textit{315}, 796-798.
\bibitem{6} Hemminger, J.; Crabtree, G. W.; Kastner, M. \emph{The Energy Challenges Report: New Science for a Secure and Sustainable Energy Future}; Argonne National Laboratory: Argonne, IL, 2008.
\bibitem{7} Man, I. C.; Su, H.; Calle-Vallejo, F.; Hansen, H. A.; Mart\1nez, J. I.; Inoglu, N. G.; Kitchin, J.; Jaramillo, T. F.; N{\o}rskov, J. K.; Rossmeisl, J. \emph{ChemCatChem} {\bf 2011} \textit{3}, 1159-1165.
\bibitem{7a} Mart\1nez, J. I.; Hansen, H. A.; Rossmeisl, J.; N{\o}rskov, J. K. \emph{Phys. Rev. B} {\bf 2009} \textit{79}, 045120(1)-045120(5).
\bibitem{8} Mowbray, D. J.; Mart\1nez, J. I.; Garc\1a-Lastra, J. M.; Thygesen, K. S.; Jacobsen, K. W. \emph{J. Phys. Chem. C} {\bf 2009} \textit{113}, 12301-12308.
\bibitem{9} Mowbray, D. J.; Mart\1nez, J. I.; Calle-Vallejo, F.; Rossmeisl, J.; Thygesen, K. S.; Jacobsen, K. W.; N{\o}rskov, J. K. \emph{J. Phys. Chem. C} {\bf 2011} \textit{115}, 2244-2252.
\bibitem{10} Chervin, C. N.; Lubers, A. M.; Pettigrew, K. A.; Long, J. W.; Westgate, M. A.; Fontanella J. J.; Rolison, D. R. \emph{Nano Lett.} {\bf 2009} \textit{9}, 2316-2321.
\bibitem{11} Krowne, C. M. \emph{Phys. Lett. A} {\bf 2010} \textit{374}, 614-619.
\bibitem{12} Krowne, C. M. \emph{Phys. Lett. A} {\bf 2010} \textit{374}, 1172-1179.
\bibitem{13} Krowne, C. M. \emph{J. Nanomater.} {\bf 2010} \textit{2010}, 160639(1)-160639(27).
\bibitem{14} An, X.; Meng, G. W.; Wei, Q.; Kong, N.; Zhang, L. D. \emph{J. Phys. Chem. B} {\bf 2006} \textit{110}, 222-226.
\bibitem{15} Li, F. J.; Zhang, S.; Kong, J. H.; Zhang, W. L. \emph{Nanosci. Nanotech. Lett.} {\bf 2011} \textit{3}, 240-245.
\bibitem{16} Bilalbegovic, G. \emph{J. Phys.: Condens. Matter} {\bf 2006} \textit{18}, 3829-3836.
\bibitem{17} Zhang, D. J.; Guo, G.; Liu, C. B.; Zhang, R. Q. \emph{J. Phys. Chem. B} {\bf 2006} \textit{110}, 23633-23636.
\bibitem{18} Lewis, J. P.; Jel\1nek, P.; Ortega, J.; Demkov, A. A.; Trabada, D. G.; Haycock, B.; Wang, H.; Adams, G.; Tomfohr, J. K.; Abad, E.; Wang, H.; Drabold, D. A. \emph{Phys. Stat. Sol. B} {\bf 2011} \textit{248}, 1989-2007.
\bibitem{19} Hammer, B.; Hansen, L. B.; N{\o}rskov, J. K. \emph{Phys. Rev. B} {\bf 1999} \textit{59}, 7413-7421.
\bibitem{20} Bahn, S. R.; Jacobsen, K. W. \emph{Comput. Sci. Eng.} {\bf 2002} \textit{4}, 56-66.
\bibitem{21} Lewis, J. P.; Glaesemann, K. R.; Voth, G. A.; Fritsch, J.; Demkov, A. A.; Ortega, J.; Sankey, O. F. \emph{Phys. Rev. B} {\bf 2001} \textit{64}, 195103(1)-195103(10).
\bibitem{22} Jel\1nek, P.; Wang, H.; Lewis, J. P.; Sankey, O. F.; Ortega, J. \emph{Phys. Rev. B} {\bf 2005} \textit{71}, 235101(1)-235101(9).
\bibitem{23} Harris, J. \emph{Phys. Rev. B} {\bf 1985} \textit{31}, 1770-1779.
\bibitem{24} Foulkes, W. M. C.; R. Haydock, \emph{Phys. Rev. B} {\bf 1989} \textit{39}, 12520-12536.
\bibitem{25} Fuchs, M.; Scheffler, M. \emph{Comput. Phys. Commun.} {\bf 1999} \textit{119}, 67-98.
\bibitem{26} Vanderbilt, D. \emph{Phys. Rev. B} \textbf{41}, 7892-7895 (1990).
\bibitem{MPG} Chadi, D. J.; Cohen, M. L. \textit{Phys. Rev. B} {\bf 1973} \textit{8}, 5747-5753.
\bibitem{27} Wyckoff, R. W. G. \emph{Crystal Structures}, Vol. 1; Interscience, John Wiley and Sons: New York, London, 1963.
\bibitem{28} Dollase, W. A. \emph{Acta Cryst.} {\bf 1967} \textit{23}, 617-623.
\bibitem{29} Kihara, K. \emph{Z. Kristallogr.} {\bf 1978} \textit{148}, 237-253.
\bibitem{30} Ross, N. L.; Shu, J. F.; Hazen, R. M.; Gasparik, T. \emph{Am. Mineralogist} {\bf 1990} \textit{75}, 739-747.
\bibitem{Alonso} Alonso, J. A. \emph{Structure and Properties of Atomic Nanoclusters}; Imperial College Press: London, 2011.
\bibitem{30a} Mart\1nez, J. I.; Castro, A.; Rubio, A.; Alonso, J. A. \emph{J. Comput. Theor. Nano.} {\bf 2006} \textit{3}, 761(1)-761(6).
\bibitem{31} Ramos, L. E.; Furthm\"uller, J.; Bechstedt, F. \emph{Phys. Rev. B} {\bf 2004} \textit{69}, 085102(1)-085102(8).
\bibitem{32} Gnani, E.; Reggiani, S.; Colle, R.; Rudan, M. \emph{IEEE Trans. Elec. Dev.} {\bf 2000} \textit{47}, 1795-1803.
\bibitem{33} Sze, S. M. \emph{Physics of Semiconductor Devices}; John Wiley and Sons: New York, 1981.
\bibitem{34} Jenichen, A.; Engler, C. \emph{Phys. Stat. Sol. B} {\bf 2010} \textit{247}, 59-66.
\bibitem{35} Picozzi, S.; Continenza, A.; Asahi, R.; Mannstadt, W.; Freeman, A. J.; Wolf, W.; Wimmer, E.; Geller, C. B. \emph{Phys. Rev. B} {\bf 2000} \textit{61}, 4677-4684.
\bibitem{36} McKelvey, J. P. \emph{Solid State and Semiconductor Physics}; Harper \& Row: New York, 1966.
\bibitem{37} Hybertsen, M. S.; Louie, S. G.; \emph{Phys. Rev. Lett.} {\bf 1985} \textit{55}, 1418-1421.
\bibitem{38} Hybertsen, M. S.; Louie, S. G.; \emph{Phys. Rev. B} {\bf 1986} \textit{34}, 5390-5413.
\bibitem{38a} Rohlfing, M.; Louie, S. G. {\it Phys. Rev. Lett.} {\bf 1999} {\it 83}, 856-859.
\bibitem{38b} Onida, G.; Reining, L.; Rubio, A.; {\it Rev. Mod. Phys.} {\bf 2002} {\it 74}, 601-659.
\bibitem{38c} Mart\1nez, J. I.; Garc\1a-Lastra, J. M.; L\'opez, M. J.; Alonso, J. A. {\it J. Chem. Phys.} {\bf 2010} {\it 132}, 044314(1)-044314(5).
\bibitem{yambo} Marini, A.; Hogan, C.; Gr\"uning, M.; Varsano, D.; {\it Comput. Phys. Commun.} {\bf 2009} {\it 180} 1392-1403.
%
\bibitem{39} Cusachs, L. C.; Politzer, P. \emph{Chem. Phys. Lett.} {\bf 1968} \textit{1}, 529-531.
\bibitem{40} L\"owdin, P. O.; \emph{Adv. Quantum Chem.} {\bf 1970} \textit{5}, 185-199.
\bibitem{41} Bader, R. F. W. \emph{Atoms in Molecules - A Quantum Theory}; Oxford University Press: New York, 1990.


\end{thebibliography}
\end{document}